\documentclass[conference, a4paper]{IEEEtran}
\usepackage[utf8]{inputenc}
\usepackage[english]{babel}
\usepackage{amsmath}
\usepackage{amssymb}
\usepackage{bbm}
\usepackage{array}
\usepackage{cmap}
\usepackage{graphicx}
\usepackage{url}
\usepackage{cite}
\usepackage{hyperref}
\usepackage{siunitx}
\usepackage{tikz}

\usetikzlibrary{arrows}
\usetikzlibrary{shapes.multipart}

\IEEEoverridecommandlockouts

\newcommand\copyrighttext{%
\footnotesize \textcopyright \enspace 2020 IEEE. Personal use of this material is permitted. Permission from IEEE must be obtained for all other uses, in any current or future media, including reprinting/republishing this material for advertising or promotional purposes, creating new collective works, for resale or redistribution to servers or lists, or reuse of any copyrighted component of this work in other works. DOI: \href{https://doi.org/10.1109/ISCC50000.2020.9219735}{10.1109/ISCC50000.2020.9219735}
}
\newcommand\copyrightnotice{%
\begin{tikzpicture}[remember picture,overlay]
\node[anchor=south] at (current page.south) {\fbox{\parbox{\dimexpr\textwidth-\fboxsep-\fboxrule\relax}{\copyrighttext}}};
\end{tikzpicture}%
}

\begin{document}
\title{An Algorithm to Satisfy the QoS Requirements\\in a Heterogeneous LoRaWAN Network\thanks{The research was done at IITP RAS and was supported by the grant No 18-37-20077 mol-a-ved of the Russian Foundation for Basic Research.}
}
\author{\IEEEauthorblockN{
Dmitry Bankov\IEEEauthorrefmark{1}\IEEEauthorrefmark{2}\IEEEauthorrefmark{3},
Evgeny Khorov\IEEEauthorrefmark{1}\IEEEauthorrefmark{2}\IEEEauthorrefmark{3}
and Andrey Lyakhov\IEEEauthorrefmark{1}\IEEEauthorrefmark{2}
}
\IEEEauthorblockA{\IEEEauthorrefmark{1}Institute for Information Transmission Problems, Russian Academy of Sciences, Moscow, Russia\\
\IEEEauthorrefmark{2}Moscow Institute of Physics and Technology, Moscow, Russia\\
\IEEEauthorrefmark{3}National Research University Higher School of Economics, Moscow, Russia\\
Email: \{bankov, khorov, lyakhov\}@iitp.ru}
}

\maketitle
\copyrightnotice

\begin{abstract}
	LoRaWAN is a popular low power wide area network technology widely used in many scenarios, such as environmental monitoring and smart cities.
	Different applications demand various quality of service (QoS), and their service within a single network requires special solutions for QoS provision.
	We consider the problem of QoS provision in heterogeneous LoRaWAN networks that consist of several groups of devices that require different packet loss rate (PLR).
	To solve this problem, we develop a mathematical model that can find the PLR distribution in a LoRaWAN network. With the model, we show that the PLR can vary significantly, and it is wrong to consider only the average PLR for the QoS provision.
	Finally, we develop an algorithm for assigning modulation and coding schemes to end-devices that provides PLRs below the required thresholds.
\end{abstract}


\section{Introduction}
LoRaWAN \cite{lorawan,bankov2016limits} is a low power wide area network technology that is gaining much popularity and is actively studied and deployed thanks to its wide coverage and low energy consumption.
LoRaWAN is used in many scenarios: environmental monitoring, e-health, smart cities, smart farming, etc.
As the popularity of LoRaWAN grows, a heterogeneous LoRaWAN infrastructure arises with multiple networks used for different applications.
A step forward would be to unite multiple LoRaWAN networks into one network, but we need solutions to satisfy different quality of service (QoS) requirements posed by various applications within a single network.

LoRaWAN uses a proprietary LoRa protocol, which provides a set of modulation and coding schemes (MCSs) with different reliability and data transmission rate.
At the MAC layer, LoRaWAN uses an open protocol \cite{lorawan}, which describes three classes of devices, the most widespread of which is class A: an ALOHA-like protocol.
With such a protocol, the network does not have many ways to influence the operation of end-devices, also called motes, and the most prominent approach to satisfy the QoS requirements of motes is to control the MCSs that they use because different LoRa MCSs are almost orthogonal to each other. The usage of different MCSs allows reducing the interference between the motes.

Most existing works on the MCS assignment in LoRaWAN networks \cite{reynders2017power, bankov2017pimrc, cuomo2017explora, qin2017resource, zorbas2018improving} focus on providing the QoS requirements, such as the packet loss rate (PLR) or throughput, for one set of motes, and optimize a single network-wide performance indicator.
Such an approach is insufficient for heterogeneous networks that consist of different groups of motes because the network-wide optimization of a parameter does not mean that the requirements of all groups are satisfied.
At the same time, the existing works on MCS assignment for heterogeneous networks \cite{bankov2019lorawan} optimize only the average values of performance indicators, but it is wrong to assign MCSs only based on the average PLR or delay since they can be distributed in a vast range within a network.
In this paper, we consider the problem of MCS assignment in a heterogeneous network to satisfy the requirements on the maximal PLR for many groups of motes.

\textit{The contribution of this paper} is threefold.
First, we develop a new model to find the PLR distribution in a LoRaWAN network. Second, we show that in a LoRaWAN network, the PLR can vary among the motes in a wide range. So many motes may have a PLR value significantly higher than the average one.
Thus, it is important to consider the maximal PLR or the PLR percentile during the MCS assignment.
Third, we develop a new MCS assignment algorithm for heterogeneous LoRaWAN networks that assigns the MCSs to the motes in such a way that for each group, the maximal value or given percentile of PLR is below the required one.

The rest of the paper is organized as follows.
Section \ref{sec:description} contains the basics of LoRaWAN channel access. Section \ref{sec:scenario} describes the scenario and the considered problem.
In Section \ref{sec:model}, we find the PLR distribution.
In Section \ref{sec:algorithm}, we construct an algorithm of MCS allocation, which ensures the limits of maximal PLR for a group of motes.
Numerical results are provided in Section \ref{sec:results}.
Conclusion is given in Section \ref{sec:conclusion}.

\section{LoRaWAN Networks}
\label{sec:description}

A LoRaWAN network typically consists of a server, gateways (GWs), and motes.
The GWs serve as relays between the server and the motes, being connected with the server via an IP network, and with motes via LoRa wireless links.

Three classes of LoRaWAN devices use different channel access rules.
We study Class A operation as the default and the most widespread one.
The network uses several wireless channels, one of which is used as a service channel, and the remaining ones are referred to as main channels.
When a mote has a data frame for transmission, it randomly selects one of the main channels and transmits the frame in it.
By default, LoRaWAN networks work in the acknowledged mode, so having received a data frame from the mote, the server responds with two acknowledgments (ACKs).
The first ACK is sent $T_1 = \SI{1}{\s}$ after the data frame reception in the main channel where the data was transmitted.
The second ACK is sent $T_2 = T_1 + \SI{1}{\s}$ after the data frame ending in the service channel.
If the mote does not receive any ACK, it makes a retransmission.
The recommended by the standard \cite{lorawan_parameters} retransmission policy is to send the frame again after a random delay from one to three seconds.

For its transmission, a mote uses an MCS $m$  assigned by the server.
The corresponding ACKs are transmitted in the main channel with MCS $m - \Delta m$, where $\Delta m$ is a configurable parameter (0 by default).
In the service channel, the ACKs are transmitted with the slowest and the most reliable MCS.

The MCSs are determined by a parameter called spreading factor (SF), and the essential feature of LoRa PHY is that the signals with different SFs are almost orthogonal to each other, i.e., a LoRa receiver can recognize two simultaneously transmitted signals with different SFs.
The orthogonality is not perfect \cite{croce2017impact}, and if the difference between the signal powers is too high, the signals can interfere with each other.
However, according to \cite{mahmood2018scalability}, in a small network with low traffic intensity, the impact of inter-SF interference is negligible.

Another essential aspect of LoRa signals is that a LoRa device can receive a signal even if it overlaps with a sufficiently weaker signal at the same MCS.
The LoRa device vendor specifies a co-channel rejection parameter of 3 dB, which is the minimum power difference required for the successful reception.
Moreover, when a device has started to receive a signal and detects another signal, it can switch to the new signal and receive it correctly, provided that its power is sufficiently higher than the power of the previous signal \cite{haxhibeqiri2017lora}.
Such behavior is called the capture effect.

The MCS assignment algorithm is not specified in the standard, which only mentions the adaptive data rate (ADR): an MCS and transmission power assignment algorithm that should make the motes use the fastest MCSs.
However, the usage of the fastest possible MCS is not the best solution in some scenarios because of the interference between devices.

A widespread approach to MCS assignment \cite{reynders2017power,  cuomo2017explora} uses the fact that the frame durations are proportional to $2^{SF}$, so the number of motes allocated to this MCS should be inversely proportional to $2^{SF}$.
With such MCS allocation, the airtime and the collision rate are almost equalized among all MCSs.

Other MCS allocation approaches involve the solution of optimization tasks.
In \cite{qin2017resource} the MCSs are allocated to maximize the minimal rate, but the network performance is calculated according to a PHY level model, which does not consider the LoRaWAN channel access rules.
In \cite{zorbas2018improving} the MCSs are assigned to maximize the number of motes for which the average success probability is higher than the given value.
This solution is based on a rather accurate mathematical model of LoRaWAN but does not take into account ACKs and retries.

The airtime equalization or optimization of a single performance indicator cannot satisfy heterogeneous requirements, and to our best knowledge, most papers about LoRaWAN do not consider the problem of MCS assignment with heterogeneous QoS requirements.
We have studied this problem in \cite{bankov2019lorawan}, where we develop a mathematical model of LoRaWAN channel access that can find the PLR and show how to use this model in order to provide the required PLR level to several groups of motes, each group demanding different PLR.
However, the developed model provides only the average PLR among all the motes in the network, while for the QoS provision the maximal PLR or the PLR percentile is much more critical, and as we show in Section \ref{sec:model}, the maximal PLR can significantly differ from the average value.

\section{Scenario and Problem Statement}
\label{sec:scenario}
We consider a LoRaWAN network with one GW and $N$ motes.
We divide the motes into $G$ groups, group $g$ having $N_g$ motes, each generating a Poisson flow of frames with rate $\lambda_g$.
The motes transmit the frames to the server via the GW, and the server acknowledges the frames with ACKs that have no payload.
As in \cite{bankov2019lorawan}, we assume that the motes have a limited buffer capacity, and if a mote generates a new frame during the transmission of another frame, the mote discards the old one after the end of the second receive window.
We also introduce a retry limit $RL$, which is a maximal number of retransmissions that a mote makes before discarding a frame.

The network operates in $F$ main channels and one downlink channel.
The motes transmit their data using the MCSs from $0$ to $M - 1$ allocated by the server.
The motes are spread uniformly around the GW within a circle with a radius $R$ that is small enough to let all the motes transmit their frames reliably with any MCS provided that no collisions occur.
We also assume that the MCSs are orthogonal.
Although this assumption is not absolutely correct, we study a small network with low network utilization, where the impact of inter-SF interference is negligible \cite{mahmood2018scalability, markkula2019simulating}.

Besides the MCSs being orthogonal, simultaneous transmissions with the same MCS  do not always lead to a collision.
We consider that a frame is received correctly if its power is at least by $Q$ dB higher than the noise and the interference at the same MCS, where $Q$ is the co-channel rejection.

The motes are divided into groups based on their requirements: for the group $g$, PLR must not exceed $PLR^{QoS}_g$.
A promising way to satisfy different QoS requirements is an appropriate allocation of MCSs to the motes because the MCSs are almost orthogonal, and the mote's PLR depends on the total load of the motes that use the same MCS.
In this paper, we design an MCS allocation algorithm guaranteeing that for each mote, the PLR does not exceed the corresponding limit. For that, we firstly analyze the drawbacks of the model from \cite{bankov2019lorawan}.
Then we develop a new model that can find the $PLR$ distribution and the maximal PLR of the motes that use a given MCS. Based on the model, we design an MCS allocation algorithm.

\section{PLR Distribution}
\label{sec:model}

\subsection{Old Approach}
The model from \cite{bankov2019lorawan} describes a single group of motes, and the PLR is calculated as follows:
\vspace{-0.3em}
\begin{equation}
\label{eq:plr_old}
\begin{split}
PLR_{old} &= 1 - \sum \limits_{i = 0}^{M - 1} p_i \Big[P_{i}^{S, 1} + \left(1 - P_{i}^{S, 1}\right) P_i^{G} \times\\
&\times \sum \limits_{r = 0}^{RL - 1} \left(P_i^{G}\left(1 - P_{i}^{S, Re}\right)\right)^{r} P_{i}^{S, Re}\Big].
\end{split}\vspace{-0.3em}
\end{equation}
Here $p_i$ is the probability of the considered mote using MCS $i$,
$P_{i}^{S, 1}$ is the probability of the first transmission attempt of the frame being successful (including the data frame and the acknowledgements transmission) when MCS $i$ is used,
$P_i^{S, Re}$ is the probability of the retransmission attempt being successful when MCS $i$ is used,
and $P_i^G$ is the probability of no new frames being generated during the transmission attempt since it is supposed that the old frame is discarded if the new one is generated.

The $P_{i}^{S, 1}$ is found as $P^{S,1}_i = P^{Data}_i P^{Ack}_{i}$, where $P^{Ack}_{i}$ is the probability of at least one ACK being delivered successfully at MCS $i$, and $P^{Data}_i$ is the successful data frame transmission probability found as
\begin{equation}
\label{eq:data1}
\begin{split}
P^{Data}_i =& e^{-(2 T^{Data}_{i} + P^{Data}_i T^{Ack}_{i}) \frac{p_i \lambda}{F}} +\\
+& \sum_{k = 1}^{N - 1} \frac{\left(2 \frac{p_i \lambda}{F} T^{Data}_{i}\right)^k}{k!} e^{-2 \frac{p_i \lambda}{F} T^{Data}_{i}} \mathbb{V}^{GW}_{i, k},
\end{split}
\end{equation}
where $T^{Data}_{i}$ and $T^{Ack}_{i}$ are the durations of data frame and ACK at MCS $i$, respectively,
$\lambda$ is the total traffic intensity for all motes in the network,
and $\mathbb{V}^{GW}_{i, k}$ is the probability of the mote's signal to exceed the power of $k$ interfering signals at least by $Q$.
Due to the space limitation, we do not provide the equations for the other successful transmission probabilities, but they are calculated similarly, and the clarification of all the equations can be found in an open-access paper \cite{bankov2019lorawan}.

\subsection{New Approach}
Note that in \cite{bankov2019lorawan}, $\mathbb{V}^{GW}_{i, k}$ and other probabilities that describe the relationship between the signal power of motes are calculated as the average values for all the motes in the network.
As a consequence, the $PLR_{old}$ is also an average value of PLR for all the motes in the network and does not show the variability of PLR for motes located at a different distance from the GW.
Let us find how the PLR of a mote depends on its distance $x$ from the GW.
To solve this problem, just as in \cite{bankov2019lorawan} we consider a log-distance path-loss model, with which the signal transmitted with the power $w_{tx}$ dBm arrives at the receiver with the power \vspace{-0.5em}
\begin{equation}
w_{rx}\left(d\right) = C_1 - C_2  \log_{10}\left(d\right),\vspace{-0.5em}%
\end{equation}where $C_1$ and $C_2$ are the constant values ($C_1$ contains $w_{tx}$ as a summand), and $d$ is the distance between the receiver and the transmitter.
As an example of such models, one can consider an empiric model that describes the LoRa links \cite{jorke2017urban}, or the Okumura-Hata model \cite{hata1980empirical}.

Let Mote 0 be at a distance $x$ from the GW and transmit a frame at MCS $i$ simultaneously with some Mote~1.
With uniform distribution of motes around the GW, the distance from Mote 1 to the GW has the following probability density function \vspace{-0.5em}
\begin{equation}
\label{eq:pdf_dist}
\rho\left(r_1\right) = \frac{2 r_1}{R^2}.\vspace{-0.5em}%
\end{equation}
In the studied scenario, there are three possible outcomes of the frame intersection.
The first one is that the GW successfully receives the frame from Mote 0 if its signal is at least by $Q$ dB more powerful than the signal from Mote 1.
Taking into account the distance distribution and the considered path-loss model, we find the probability of such an event as follows:
\begin{equation}
\begin{split}
&\mathbb{V}^{GW}_{i, 1}\left(x\right) = \mathbb{P} \left(w^{G}(r_1) < w^{G}(x) - Q\right) = \\
&= \mathbb{P} \left(C_1 - C_2 \lg(r_1) < C_1 - C_2 \lg(x) - Q\right) = \\
&= \mathbb{P} \left(r_1 > x 10^{\frac{Q}{C_2}}\right) = \int\limits_{min(R, x \cdot 10^{\frac{Q}{C_2}})}^R \frac{2 r_1}{R^2} d r_1 = \\
&= \begin{cases}
0, & x > R \cdot 10^{-\frac{Q}{C_2}}, \\
1 - \frac{x^2 \cdot 10^{\frac{2 Q}{C_2}}}{R^2}, & x \leq R \cdot 10^{-\frac{Q}{C_2}}.
\end{cases}
\end{split}
\end{equation}
Here $w^G(x)$ is the power of a Mote's signal at the GW if the Mote's distance from the GW is $x$.

The second possible outcome is that both the intersecting frames are damaged, which happens when the difference of the frames' powers is less than $Q$.
The probability of such an outcome is found in a similar way:
\begin{equation}
\begin{split}
&\mathbb{V}^{Both}_i\left(x\right) = \int\limits_{x \cdot 10^{-\frac{Q}{C_2}}}^{min(R, x \cdot 10^{\frac{Q}{C_2}})} \frac{2 r_1}{R^2} d r_1 = \\
=& \begin{cases}
1 - \frac{x^2 \cdot 10^{-\frac{2 Q}{C_2}}}{R^2}, & x > R \cdot 10^{-\frac{Q}{C_2}}, \\
\frac{x^2}{R^2} \left(10^{\frac{2Q}{C_2}} - 10^{-\frac{2Q}{C_2}}\right), & x \leq R \cdot 10^{-\frac{Q}{C_2}}.
\end{cases}
\end{split}
\end{equation}

The third possible outcome is that the frame of Mote 0 is not received, while the frame of Mote 1 is received successfully.
The probability of this outcome is
\begin{equation}
\mathbb{V}^{One}_i(x) = 1 - \mathbb{V}^{GW}_{i, 1}(x) - \mathbb{V}^{Both}_i(x) = \frac{x^2 \cdot 10^{-2\frac{Q}{C_2}}}{R^2}.
\end{equation}

It is essential to differentiate between the mentioned three possibilities, because in the first case Mote 0 makes a successful transmission, in the second case there is a collision and both Mote 0 and Mote 1 retransmits, while in the third case only Mote 0 retransmits and the probability of successful retransmission differs for these cases.

Let us now consider that Mote 0 has made a successful data frame transmission; the GW responds with an ACK at MCS $i$, and this transmission interferes with a frame of Mote 1.
The ACK is delivered successfully if at Mote 0 the signal $w^{M}_0(x)$ from the GW is stronger than the signal $w^{M}_1$ from Mote 1 by at least $Q$ dB.
Such an event happens with probability
\begin{equation}
\begin{split}
&\mathbb{V}^{Mote}_{i, 1}(x) = \mathbb{P} \left(w^{M}_1 < w^{M}_0(x) - Q\right)= \\
&= \mathbb{P} \left(d_1 > x 10^{\frac{Q}{C_2}}\right) = \iint\limits_{\mathcal{A}(x)} \frac{2 r_1}{R^2} \frac{d\phi} {2 \pi} d r_1 =\\
=& \int\limits_{0}^{R}\frac{r_1}{R^2} \left(2\arccos\left(\frac{x^2 + r_1^2 - x^2 10^{\frac{2Q}{C_2}}}{2 x r_1}\right) - \pi\right) d r_1,
\end{split}
\end{equation}
where $d_1$ is the distance from Mote 1 to Mote 0, and $\mathcal{A}(x)$ is the domain of such polar coordinates $(r_1, \phi)$ of Mote 1 that the power condition holds:
\[
\resizebox{\linewidth}{!}{
	$\mathcal{A}(x) = \left\{r_1, \phi: 0 < r_1 < R \wedge cos \phi \leq \frac{x^2 + r_1^2 - x^2 10^{\frac{2Q}{C_2}}}{2 x r_1}\right\}$.
}
\]

The probabilities $\mathbb{V}^{*}_{*}(x)$ for a given distance $x$ of the transmitting mote describe possible outcomes of its transmission in the presence of interfering frame.
The values $P_{i}^{S, 1}$ and $P_i^{S, Re}$ in the equation \eqref{eq:plr_old} contain similar values, but in \cite{bankov2019lorawan} these values are found as an average for all the possible Mote 0 locations within the circe and do not depend on $x$.
We propose to replace these values with $\mathbb{V}^{*}_{*}(x)$ because they contain the dependency on $x$.
We also take into account multiple groups of devices which generate different network load.
Let $\vec{a}_i = \left[a_{i, 0}, a_{i, 1}, ..., a_{i, G - 1}\right]$ be the allocation vector which contains the numbers of motes from different groups assigned to the MCS $i$.
They generate $l_i = \sum_g a_{i, g} \lambda_g$ frames per second in total.
With such a new notation, we change the $P^{Data}_i$ as follows
\begin{equation}
\label{eq:data_new}
\begin{split}
P^{Data}_{i, g}(x) =& e^{-(2 T^{Data}_{i} + P^{Data}_{i, g}(x) T^{Ack}_{i}) \frac{l_i - \lambda_g}{F}} +\\
+& 2 \frac{l_i - \lambda_g}{F} T^{Data}_{i} e^{-2 \frac{l_i - \lambda_g}{F} T^{Data}_{i}} \mathbb{V}^{GW}_{i, 1}(x).
\end{split}
\end{equation}
Here we added the dependency on the mote distance $x$ and the group number $g$, and changed the network load at MCS $i$ from $p_i \lambda$ to $l_i - \lambda_g$, i.e., the traffic generation rate taken for all the motes except the considered one.
Note that we leave only two summands from \eqref{eq:data1} and do not consider the interference between more than two motes.
The same simplification has been used in \cite{bankov2019lorawan}, where it is shown that it does not significantly affect the accuracy of the model.

The PLR for a mote from group $g$ that transmits with MCS $i$ and is located at a distance $x$ from the GW is found as
\begin{equation}
\label{eq:plr}
\begin{split}
PLR_{i, g}(x) &= 1 - P_{i, g}^{S, 1}(x) - \left(1 - P_{i, g}^{S, 1}(x)\right) P_{i, g}^{G} \times \\
\times & \sum \limits_{r = 0}^{RL - 1} \left[P_{i, g}^{G}\left(1 - P_{i, g}^{S, Re}(x)\right)\right]^{r} P_{i, g}^{S, Re}(x),
\end{split}
\end{equation}
where $P_{i, g}^{S, 1}$ and $P_{i, g}^{S, Re}$ now depend on $x$, because the probabilities of power relations now depend on $x$, and all the probabilities in the equation depend on the group.
The meaning of this equation is that the mote loses the frame if it has a transmission failure after the first transmission attempt, and during the subsequent retransmissions, it either discards the frame due to a new arriving frame or makes $RL$ unsuccessful retransmission attempts.

With $\rho(r)$ and $PLR_{i, g}(x)$ we find the PLR distribution:
\begin{equation}
\mathbb{P}(PLR_{i, g} < y) = \int\limits_{0}^{R} \mathbb{I}\left(PLR_{i, g}(r) < y\right)\rho(r) d r,
\end{equation}
where $\mathbb{I}(x)$ equals 1 if $x$ is true and 0, otherwise.
In Section~\ref{sec:results}, we show that the PLR distribution is concentrated around the maximal PLR value, and it is important to consider the maximal value or the percentile of the PLR while assigning the MCSs to the motes.

\section{MCS Allocation Algorithm}
\label{sec:algorithm}
Let us now design an heuristic MCS allocation algorithm that uses a greedy approach to solve the problem stated in Section \ref{sec:scenario}.
Our algorithm uses the developed model to find the maximal PLR for motes that generate a given traffic intensity at MCS $i$:
\begin{equation}
\pi_{i, g}(\lambda, \lambda_g, l_i) \triangleq \max_{x \in [0, R]} PLR_{i_g}(x),
\end{equation}
where we explicitly indicate that this value depends on the total network load, network load for mote from group $g$ and the total load from motes assigned to MCS $i$.

If motes from group $g$ are allocated MCS $i$, then the algorithm should guarantee that $\pi_{i, g}(\lambda, \lambda_g, l_i)$ is less than $PLR^{QoS}_g$.
Moreover, if motes from several groups use the same MCS, this inequality should hold for all the groups.

To solve this problem, we define an auxiliary value $\nu^g_i$ as the maximal network load from group $g$ motes which can use MCS $i$ to yield the PLR less than $PLR^{QoS}_g$, provided that other motes do not use this MCS:
\begin{equation}
\nu^g_i \triangleq \max\left\{l: \pi_{i, g}\left(\lambda, \lambda_g, l\right) \leq PLR^{QoS}_g\right\}.
\end{equation}
This value can be interpreted as the network capacity for a given group at a given MCS.

\begin{figure}[tb]
	\center{\includegraphics[width=0.8\linewidth]{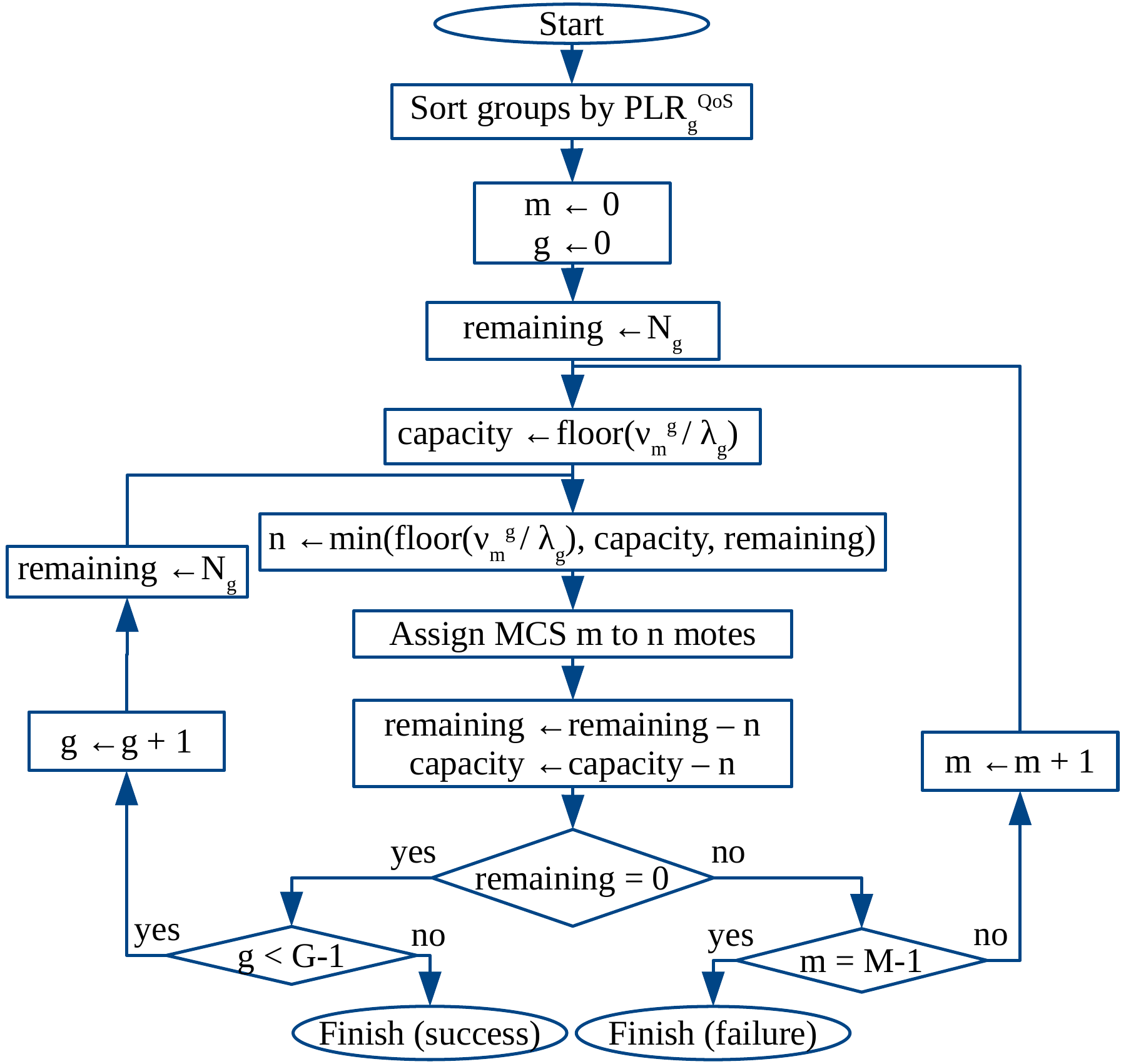}}
	\vspace{-1em}
	\caption{Scheme of the MCS Allocation Algorithm.}
	\label{fig:scheme}
	\vspace{-1.5em}
\end{figure}

With this value, we construct the following MCS allocation algorithm.
Its scheme is shown in Fig. \ref{fig:scheme}, where ''$\leftarrow$`` is the assignment operator.
Firstly, the server sorts the groups by $\nu^g_i$ in the ascending order.
Then it starts with MCS 0 and group~0: the slowest MCS with the lowest capacity and the group with the strictest PLR requirement.
The server allocates MCS 0 to a minimum between $N_0$ and $\lfloor\frac{\nu^0_0}{\lambda_0}\rfloor$ motes from group 1: it cannot allocate MCS 1 to more motes, because either the PLR requirement will not be satisfied, or there are no more motes from group 0 without an assigned MCS.
If $\nu^0_0$ is less than the number $N_0$ of motes from group 0, the algorithm transits to MCS 1 and assigns MCS 1 to the minimum between $N_0 - \lfloor\frac{\nu^0_0}{\lambda_0}\rfloor$ and $\lfloor\frac{\nu^0_1}{\lambda_0}\rfloor$ motes from group 0.
Such a procedure is done until all the motes from group 0 have an assigned MCS.

Let the procedure stop at MCS $m$, which was allocated to $n$ motes from group 0.
The server transits to group 1 and tries to allocate MCS $m$ to motes from this group.
However, it should guarantee that the PLR requirement is fulfilled both for group 0 and group 1.
So the server assigns MCS $m$ to a number of motes from group 1, equal to the minimum between $\lfloor\frac{\nu^1_m}{\lambda_1}\rfloor$ and $\nu^0_m - n$.
If not all the motes from group 1 have an MCS, then the server transits to the next MCS and continues the previously described procedure.
In the same way, the server assigns MCSs to motes from the remaining groups.

If the algorithm exhausts the capacity of all MCSs, but still some motes are assigned to an MCS, then the algorithm reports a failure.
Otherwise, the algorithm is completed successfully.

\let\times\cdot
\section{Numerical Results}
\label{sec:results}
\subsection{Packet Loss Rate}
\begin{figure}[tb]
	\vspace{-1em}
	\center{\includegraphics[width=0.85\linewidth]{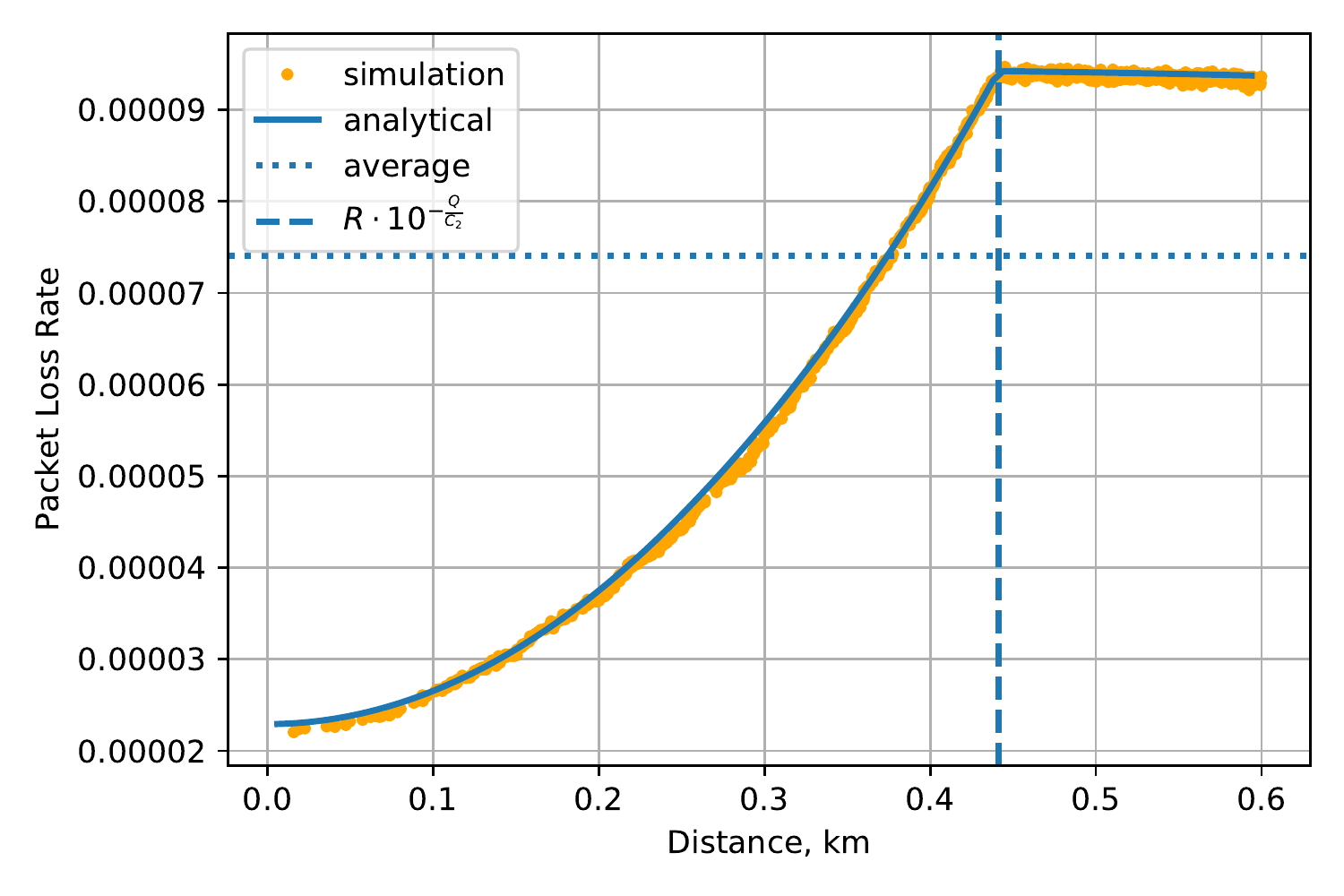}}
	\vspace{-1em}
	\caption{Dependency of PLR on the mote's distance from the GW.}
	\label{fig:plr_dist}
	\vspace{-1em}
\end{figure}

Let us now study the distribution of PLR within the circle.
To illustrate the inequality of PLR for motes at a different distance from the GW, we consider a scenario with $1000$ motes that generate a total flow of $0.5$ frames per second, use only MCS $5$ and are spread within a circle with radius \SI{600}{\m}.
The co-channel rejection parameter $Q$ equals $6$ dB, the retry limit $RL$ is $7$, and the path-loss parameters $C_1$ and $C_2$ equal $-133.7$ dBm, and $44.9$ dB, respectively.
Fig. \ref{fig:plr_dist} shows the dependency of PLR on the distance of a mote to the GW obtained with \eqref{eq:plr} and with simulation.
We note that our model provides very accurate results.
As one can see, the capture effect creates inequality in PLR: motes located close to the GW have lower PLR because they have a high probability of delivering their frames regardless of the interference.
The PLR grows with $x$, and at $x = R\cdot 10^{-\frac{Q}{C_2}}$ reaches a peak, after which the PLR slightly decreases.
The motes which are far from the GW have almost the same PLR, close to its maximal value.
The peak location depends on the co-channel rejection parameter $Q$, while without capture effect, i.e., with $Q \rightarrow \infty$, there is no peak at all, and the PLR is constant for all motes.
It is important to note that the average PLR, calculated according to \eqref{eq:plr_old}, differs from the maximal PLR by almost 30\%.

\begin{figure}[tb]
	\center{\includegraphics[width=0.85\linewidth]{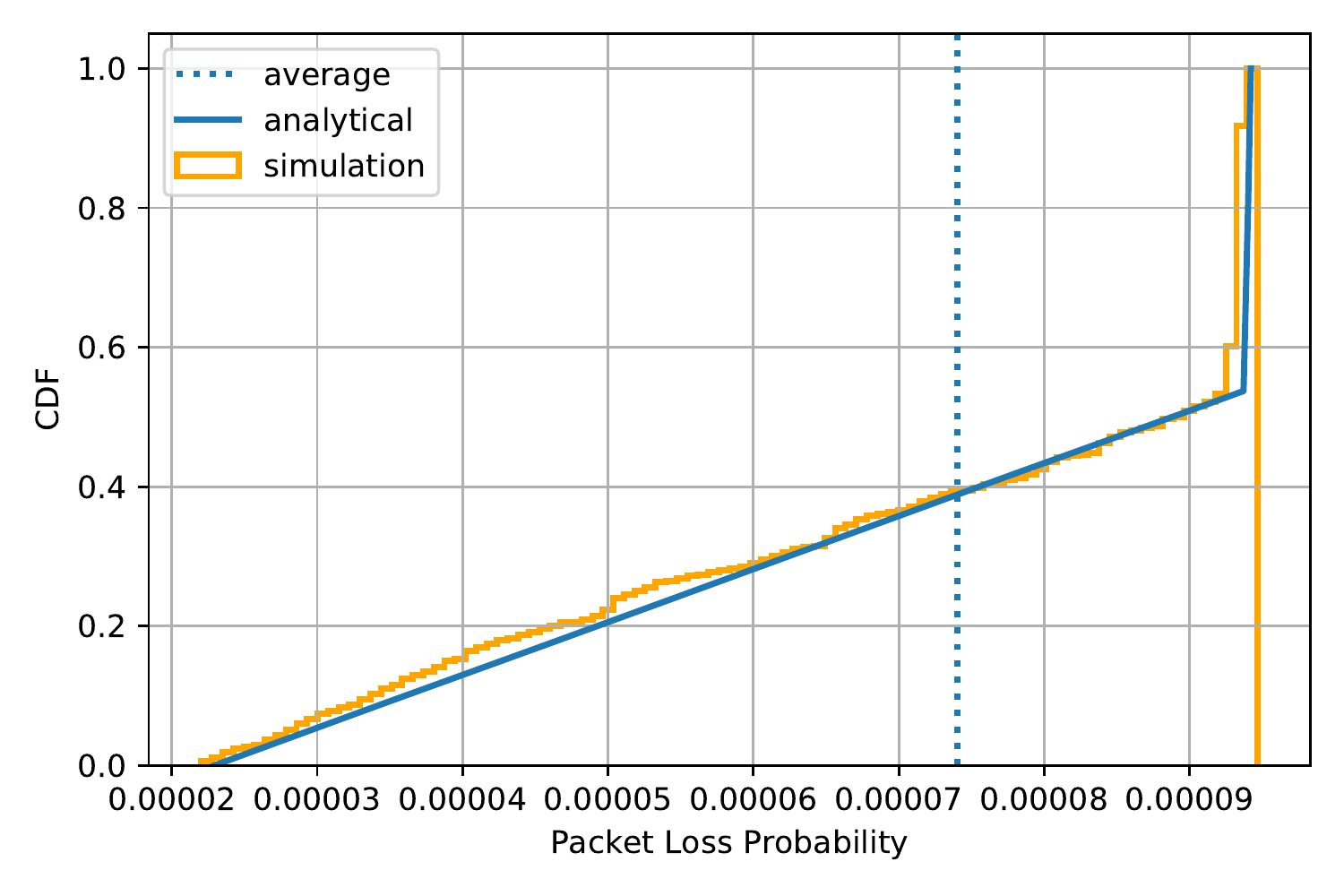}}
	\vspace{-1em}
	\caption{Cumulative distribution function of PLR.}
	\label{fig:plr_cdf}
	\vspace{-1em}
\end{figure}

Let us now consider the PLR distribution.
The motes are distributed uniformly within the circle, their distance from the GW has a pdf $\rho(r)$ given in \eqref{eq:pdf_dist}, i.e., the number of motes at a distance $x$ grows linearly with the distance,
The linear growth of the mote density and of the PLR with the distance results in almost 50\% of the motes having the maximal PLR, as shown in Fig. \ref{fig:plr_cdf}, so a majority of motes have PLR, which is 30\% higher than the average value!

So, the MCS allocation for motes that require limited PLR should be made, taking into account the maximal PLR, not the average value.

\subsection{Algorithm Operation}
To illustrate the algorithm operation, we consider a network with three groups of motes:
\begin{itemize}
	\item Group $0$ consists of 10 motes each generating 0.0001 frames per second and require PLR less than $10^{-7}$,
	\item Group $1$ consists of 100 motes each generating 0.0001 frames per second and require PLR less than $10^{-6}$,
	\item Group $2$ consists of 1000 motes, each generating 0.0001 frames per second and requires PLR less than $10^{-5}$.
\end{itemize}

The PLR requirements of the groups define their $\nu^g_i$ capacity values for all MCSs, shown in Table~\ref{tab:capacities}.
Given these values, the server uses the algorithm to find an MCS allocation that satisfies the PLR requirements.
The server firstly considers Group~0 and assigns MCSs 0, 1, and 2 to the maximal possible number of motes, i.e., to one, two, and four motes, respectively.
The remaining three motes from Group~0 use MCS~2, but the capacity of this MCS is not exhausted, so the server transits to Group~1 and assigns MCS~4 to additional $7 - 3 = 4$ motes.
At this step, the number of motes which can use MCS~3 is limited by the capacity $\nu^0_3$ for the Group~1, so even though the capacity $\nu^1_3$ for Group~1 is higher, the server cannot assign MCS to more motes from Group~1, because the PLR requirement for motes from Group~0 will not be satisfied.
MCS~4 has enough capacity for the remaining 96 motes from Group~1, and it also has the capacity for $132 - 96 = 36$ motes from Group~2.
The remaining motes from Group~2 use MCS~5.
The resulting MCS assignment is shown in Table \ref{tab:assignment}.
As one can see, with such an MCS assignment, the PLR for motes that use an MCS is less than their strictest $PLR^{QoS}$ requirement.

\begin{table}[tb]
	\vspace{-1em}
	\begin{center}
		\caption{\label{tab:capacities} Network capacity for various mote groups and MCSs.}
		\vspace{-1em}
		\begin{tabular}{|c|ccc|} \hline
			& \multicolumn{3}{c|}{\textbf{Capacity, s$^{-1}$}} \\ \hline
			\textbf{MCS} & \textbf{Group 0} & \textbf{Group 1} & \textbf{Group 2} \\ \hline
			0 & 0.0001 & 0.0006 & 0.006 \\ 
			1 & 0.0002 & 0.0014 & 0.014 \\ 
			2 & 0.0004 & 0.0034 & 0.034 \\ 
			3 & 0.0007 & 0.0069 & 0.069 \\ 
			4 & 0.0014 & 0.0132 & 0.133 \\ 
			5 & 0.0026 & 0.0255 & 0.263 \\ \hline
		\end{tabular}
	\end{center}
	\vspace{-1em}
\end{table}

\begin{table}[tb]
	\begin{center}
		\vspace{-1em}
		\caption{\label{tab:assignment} MCS assignment that satisfies the QoS requirements.}
		\vspace{-1em}
		\begin{tabular}{|c|ccc|c|} \hline
			\textbf{MCS} & \textbf{Group 0} & \textbf{Group 1} & \textbf{Group 2} & \textbf{Max. PLR} \\ \hline
			0 & 1 & 0  & 0   & 0 \\ 
			1 & 2 & 0  & 0   & $7.2 \times 10^{-8}$ \\ 
			2 & 4 & 0  & 0   & $8.9 \times 10^{-8}$ \\ 
			3 & 3 & 4  & 0   & $8.8 \times 10^{-8}$ \\ 
			4 & 0 & 96 & 36  & $9.9 \times 10^{-7}$ \\ 
			5 & 0 & 0  & 964 & $3.8 \times 10^{-6}$ \\ \hline
			\textbf{Req. PLR} & $10^{-7}$ & $10^{-6}$  & $10^{-5}$ & \\
			\hline
			{\textbf{Max. PLR}} & $8.9 \times 10^{-8}$ & $9.9 \times 10^{-7}$  & $3.8 \times 10^{-6}$ & \\
			\hline
		\end{tabular}
	\end{center}
	\vspace{-2em}
\end{table}

If we change the $\pi_{i, g}$ in the algorithm to the average PLR given in \eqref{eq:plr_old}, the algorithm will assign MCSs as shown in Table~\ref{tab:assignment_average}.
Although the average PLR equals $PLR^{QoS}_g$, for some motes the PLR can still exceed the $PLR^{QoS}_g$, e.g., for MCS 2 the maximal PLR equals $1.2 \times 10^{-7}$, while devices from Group 0 require PLR less than $10^{-7}$.
The PLR requirement is also not met for devices using MCS 4 from Group 2.
Thus we confirm that the MCS allocation should be made while taking into account the maximal PLR, but not the average one.

\begin{table}[tb]\vspace{-1em}
	\begin{center}
		\caption{\label{tab:assignment_average} MCS assignment that considers only the average PLR.}
		\vspace{-1em}
		\begin{tabular}{|c|ccc|c|} \hline
			\textbf{MCS} & \textbf{Group 0} & \textbf{Group 1} & \textbf{Group 2} & \textbf{Max. PLR} \\ \hline
			0 & 1 & 0  & 0   & 0 \\ 
			1 & 2 & 0  & 0   & $7.2 \times 10^{-8}$ \\ 
			2 & 5 & 0  & 0   & $1.2 \times 10^{-7}$ \\ 
			3 & 2 & 7  & 0   & $1.2 \times 10^{-7}$ \\ 
			4 & 0 & 93 & 76  & $1.3 \times 10^{-6}$ \\ 
			5 & 0 & 0  & 924 & $3.6 \times 10^{-6}$ \\ \hline
			\textbf{Req. PLR} \ & $10^{-7}$ & $10^{-6}$  & $10^{-5}$ &\\
			\hline
			\textbf{Max. PLR}\  & $1.2 \times 10^{-7}$ & $1.3 \times 10^{-6}$  & $3.6 \times 10^{-6}$ &\\
			\hline
		\end{tabular}
		\vspace{-1em}
	\end{center}
	\vspace{-1em}
\end{table}

Let us also show an example when the algorithm cannot allocate the MCSs to satisfy the PLR requirements for all motes.
Consider the same three groups, but Group 0 has 20 motes, each generating 0.0001 frames per second.
In such a case, the total capacity of MCS 0, ..., MCS 3 is not sufficient, and the server has to allocate MCS 4 to the remaining six motes from Group 0.
When the server further considers Group 1, the remaining capacity of MCS 4 is only $14 - 6 = 8$ motes, so the server assigns MCS 4 to these motes, while MCS 5 is allocated to the remaining 94 motes from Group 1.
The remaining capacity of MCS 5 is $255 - 94 = 161$, which is not enough for Group 2, so the algorithm signalizes the allocation failure.
The allocation failure indicates that it is needed to use some admission control methods to limit the number of motes in the network, or more GWs to extend the network capacity.

%

\section{Conclusion}
\label{sec:conclusion}
In this paper, we have studied the problem of MCS assignment in a LoRaWAN network, where devices have different QoS requirements, e.g., set different limits on the PLR.
We have developed a new model to find the PLR distribution and have shown that in a LoRaWAN network, it is a possible situation that, for a majority of motes, the PLR exceeds the average value by 30\%.
Such a PLR distribution means that the provision of the required PLR should be done, taking into account the maximal PLR or the PLR percentile, while the previous approaches considering only the average PLR are erroneous.

We have designed an MCS allocation algorithm which considers several groups of motes with different PLR requirements and traffic intensity and assigns them MCSs in such a way, that the maximal PLR or the PLR percentile do not exceed the given values for all the motes.

\bibliographystyle{IEEEtran}
\bibliography{biblio}

\end{document}